\documentclass[11pt]{article}

\usepackage{epsfig}
\usepackage{times}
\usepackage{epstopdf}
\usepackage{amsmath}
\usepackage{amssymb}
\usepackage[left=1.75cm,right=1.75cm,top=3.0cm,bottom=3.0cm]{geometry}
\usepackage{setspace}

\begin{document}

\bigskip\bigskip
\centerline {\Large \bf {Quasilocal Conservation Laws: Why We Need Them}}
\bigskip\bigskip
\centerline{\large Paul L. McGrath$^{a}$, Richard J. Epp$^{b}$ and Robert B. Mann$^{a,b}$}
\bigskip\bigskip
\centerline{${}^a$ \em Department of Physics and Astronomy, University of Waterloo, Waterloo, Ontario N2L 3G1, Canada}
\vspace{0.2 cm}
\centerline{${}^b$ \em Perimeter Institute for Theoretical Physics, Waterloo, Ontario N2L 2Y5, Canada}
\vspace{0.3cm}
\centerline{\em  pmcgrath@uwaterloo.ca, repp@perimeterinstitute.ca, rbmann@sciborg.uwaterloo.ca}
\bigskip\bigskip

\begin{abstract}
We argue that conservation laws based on the {\it local} matter-only stress-energy-momentum tensor (characterized by energy and momentum per unit {\it volume}) cannot adequately explain a wide variety of even very simple physical phenomena because they fail to properly account for gravitational effects. We construct a general {\it quasi}local conservation law based on the Brown and York {\it total} (matter {\it plus} gravity) stress-energy-momentum tensor (characterized by energy and momentum per unit {\it area}), and argue that it {\it does} properly account for gravitational effects. As a simple example of the explanatory power of this quasilocal approach, consider that, when we accelerate toward a freely-floating massive object, the kinetic energy of that object increases (relative to our frame). But how, exactly, does the object acquire this increasing kinetic energy? Using the energy form of our quasilocal conservation law, we can see precisely the actual mechanism by which the kinetic energy increases: It is due to a bona fide gravitational energy flux that is exactly analogous to the electromagnetic Poynting flux, and involves the general relativistic effect of frame dragging caused by the object's motion relative to us.
\end{abstract}

\onehalfspacing

\section{Introduction and Summary} \label{Introduction}

A conservation law ought to explain the change in some physical quantity contained inside a volume of space (e.g., total energy) in terms of related fluxes passing through the bounding surface of that volume. The standard approach to constructing conservation laws is based on the identity: $\nabla_a ( T^{ab}\Psi_b ) = ( {\nabla_a T^{ab}} ) \Psi_b + T^{ab} \nabla_{(a} \Psi_{b)}$, where $T^{ab}$ is the matter stress-energy-momentum tensor, and $\Psi^a$ is a vector that determines the type of conservation law, viz., energy, momentum, or angular momentum. In the context of general relativity, matter energy-momentum is locally covariantly conserved, i.e., $\nabla_a T^{ab}=0$, and the identity reduces to $\nabla_a ( T^{ab}\Psi_b ) = T^{ab} \nabla_{(a} \Psi_{b)}$. As is well known, the problem with this local conservation law is that the right hand side is, in general, not zero (or even a covariant divergence), resulting in a bulk term in the integrated conservation law that spoils what a conservation law ought to be.

This problem is essentially gravitational in nature. There are several ways to see this: (1) The offending bulk term disappears when the spacetime has a suitable symmetry, i.e., admits a Killing vector, $\Psi^a$, but dynamically interesting spacetimes (e.g., ones containing gravitational effects due to objects in motion) generically do not. The idea of relying on a spacetime symmetry to construct a conservation law is a throwback to pre-general relativity days, e.g., special relativity, where spacetime is maximally symmetric. The same goes for relying on asymptotic spacetime symmetries, where we are still in essentially a special relativistic mindset. Given that gravity {\it is} nontrivial spacetime geometry, an approach relying on spacetime symmetries cannot hope to properly incorporate gravitational effects in general. (2) The local conservation law above is {\it homogeneous} in $T^{ab}$. In any matter-free region it is vacuous, even if that region contains interesting gravitational physics, e.g., gravitational waves. The local conservation law is essentially blind to gravitational physics. We need a conservation law that is nontrivial even when $T^{ab}=0$. (3) Because of the equivalence principle, gravitational effects, e.g., gravitational energy, are not localizable, so we have no hope of capturing gravitational physics with a conservation law based on a local stress-energy-momentum tensor. For example, there is no such thing as a local gravitational energy density (energy per unit volume), that when integrated over a volume gives the total gravitational energy in that volume (see, e.g., {\S}20.4 of reference \cite{MTW}). (4) A bulk term in a local conservation law can be a symptom of the presence of fields that are not being accounted for in the stress-energy-momentum tensor. For example, in the standard Poynting theorem, the $\vec{j}\cdot\vec{E}$ bulk term is present because $T^{ab}$ excludes the charged matter field that is the source of the electromagnetic field, and represents an energy transfer mechanism between the electromagnetic field and the charged matter field. We contend that the bulk term in the local conservation law is, similarly, a result of $T^{ab}$ not properly accounting for the physics of the gravitational field.

A solution to this problem is to move from local to {\it quasi}local conservation laws, which {\it can} properly account for the gravitational physics. In this paper we construct a general quasilocal conservation law based not on the local {\it matter} stress-energy-momentum tensor, but on the Brown and York quasilocal {\it total} stress-energy-momentum tensor (matter {\it plus} gravity) \cite{BY1993}. Here, ``quasilocal" means that the differential conservation law is integrated not over the history of a volume of space, but over the history of the {\it boundary} of that volume. We focus on the case of energy conservation, and show that in the quasilocal approach, the quasilocal analogue of the offending $T^{ab} \nabla_{(a} \Psi_{b)}$ bulk term becomes a {\it surface flux} term, which immediately solves the main problem mentioned in the opening paragraph above. Moreover, this surface flux term has two components: (I) The first component is a ``stress times strain" term that can always be made to vanish by a suitable choice of frame, called a {\it Rigid Quasilocal Frame} (or RQF) \cite{EMM2009, EMM2011}. The reader need not be familiar with RQFs to read this paper; it is sufficient to mention that, regarding point (1) in the previous paragraph, an RQF satisfies a certain ``quasilocally projected" form of the timelike Killing vector condition for stationary spacetimes that allows us to move just far enough away from the spacetime symmetry mindset to include generic (i.e., non-stationary) spacetimes in conservation laws for energy, momentum, and angular momentum. (II) The second component is an ``acceleration times momentum" term. This term is familiar from classical mechanics, and represents the rate at which the kinetic energy of an object increases due to one's acceleration toward it. Motivated by a simple equivalence principle argument, we show that this second term is actually a {\it gravitational} energy flux involving the general relativistic effect of frame dragging. We thus show precisely how quasilocal conservation laws resolve the bulk term problem in local conservation laws by properly accounting for the physics of the gravitational field.

Our paper is organized as follows. In {\S}\ref{Paradox} we introduce a very simple example of energy conservation in the context of special relativity, for the purpose of having a concrete example with which to illustrate the development of the general ideas. We consider a variant of Bell's spaceship paradox in which a box accelerates rigidly  in a transverse, uniform electric field. Obviously, the electromagnetic energy inside the box increases, but how would co-moving observers explain this increase? Paradoxically, only {\it half} of the increasing energy comes from a net Poynting flux, and, according to the local energy conservation law, the other half comes from a bulk ``acceleration times momentum" term integrated over the volume of the box. In {\S}\ref{Local} we examine local conservation laws in general, with a particular focus on the role played by the $T^{ab} \nabla_{(a} \Psi_{b)}$ bulk term in our paradox example. This provides a point of comparison for {\S}\ref{Quasilocal}, in which we construct a general quasilocal conservation law and argue how it properly accounts for gravitational physics. We also apply the energy form of this quasilocal conservation law to the general relativistic version of our paradox example to concretely illustrate how, what we would normally think of as a bulk ``acceleration times momentum" term, is actually a gravitational energy flux entering through the boundary of the box. In {\S}\ref{Conclusions} we present a complementary summary and conclusions, and argue that quasilocal conservation laws are necessary to understand more deeply a wide variety of phenomena, including the simple example of dropping an apple.

\section{An Apparent Paradox and its Resolution}\label{Paradox}

\subsection*{Paradox Outline}

\begin{figure}
\begin{center}
\includegraphics[scale=1.2]{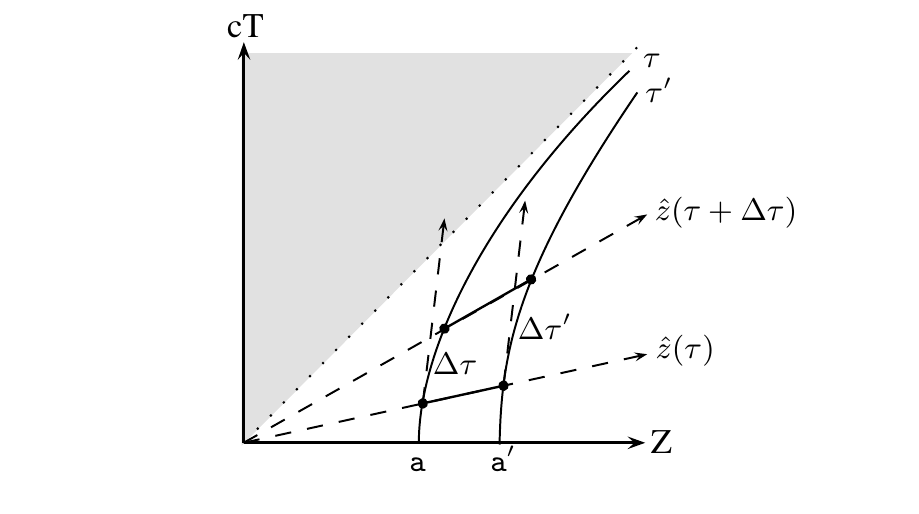}\label{trajectory}
\caption{The trajectory of a rigidly accelerating cylinder}
\end{center}
\end{figure}

Consider a right cylinder of length $L$ and cross-sectional area $A$, whose axis is parallel to the $Z$-axis of an inertial reference frame in flat spacetime with Minkowski coordinates $(cT,X,Y,Z)$. The cylinder is immersed in a constant, uniform electric field of magnitude $E$ in the positive $X$-direction, and thus contains an electromagnetic energy density equal to $E^2/8\pi$. Now let us give the cylinder constant proper acceleration in the positive $Z$-direction so that the length (and volume) of the cylinder remain fixed for co-moving observers.  It is well-known from Bell's spaceship paradox that, in order for the length of the cylinder to remain constant for co-moving observers, the top of the cylinder (represented by the hyperbola on the right in the diagram) must experience {\it less} proper acceleration, namely $\mathtt{a}^\prime=\mathtt{a}/(1+\mathtt{a}L/c^2)$, compared to $\mathtt{a}$, the proper acceleration experienced at the bottom of the cylinder (the hyperbola on the left) \cite{Bell}. The following two facts, also illustrated in the diagram, will be important to us: (1) The straight lines passing through the origin represent a natural choice for the surfaces of simultaneity for the co-moving observers: on any such simultaneity, all of the accelerating co-moving observers see the {\it same} instantaneous velocity, $v$, relative to observers at rest in the inertial frame. In other words, this is a ``constant $v$" time foliation, with $v$ monotonically increasing with time. (2) Co-moving observers at different positions along the length of the cylinder will see proper time flowing at different rates relative to one another. That is, between two co-moving simultaneities, if a proper time $\Delta\tau$ elapses for an observer at the bottom of the cylinder, a {\it greater} proper time, $\Delta\tau^\prime =(1+\mathtt{a}L/c^2)\Delta\tau$ elapses for an observer at the top.

With these facts in mind, let us consider the proper electric and magnetic fields seen by the co-moving observers. Let $\tau$ (respectively, $\tau^\prime$) denote the proper time of observers at the bottom (respectively, top) of the cylinder, and $(\hat{x},\hat{y},\hat{z})$ denote the natural choice of co-moving spatial Cartesian unit vectors. Since the observers are moving in a direction perpendicular to the electric field, they will see, in addition to a stronger electric field, a magnetic field; specifically, $\vec{E} = \hat{x} \gamma E $ and $\vec{B} = - \hat{y}\beta \gamma E $, where $\beta=v/c$ and $\gamma=1/\sqrt{1-\beta^2}$. It is important to note that since these fields depend only on $v$, all observers on any given co-moving simultaneity will see the {\it same} instantaneous electric and magnetic fields. They will thus see the {\it same} proper Poynting vector, $\vec{S} = \frac{c}{4\pi}\vec{E} \times \vec{B} = - \hat{z}\frac{c}{4 \pi} \beta \gamma^2 E^2$, and the {\it same} proper electromagnetic energy density, $u = \frac{1}{8 \pi} (|\vec{E}|^{2} + |\vec{B}|^{2}) = \frac{1}{8 \pi} (1 + \beta^2) \gamma^2 E^2$. Notice that, according to this expression for $u$, the total electromagnetic energy inside the cylinder is clearly increasing with time. We ask: How would the co-moving observers explain this increasing electromagnetic energy? One might expect that the change in energy between two co-moving simultaneities is just equal to the net Poynting flux into the cylinder over that time interval. As we will see, this is only half correct.

Consider a pair of infinitesimally separated co-moving simultaneities labeled by proper time $\tau$ and $\tau+\Delta\tau$ as experienced by observers at the bottom of the cylinder. (A bit of thought shows that it does not matter whose proper time we use to parameterize the simultaneities.)  Between these two simultaneities the total electromagnetic energy inside the cylinder changes by an amount
\begin{eqnarray} \label{E(1)}
\Delta E= \frac{du}{d\tau} \Delta \tau = 4 \beta\gamma^2 \left( \frac{E^2}{8\pi } V \right)   \frac{\mathtt{a}\Delta\tau}{c},
\end{eqnarray}
where $V=AL$ is the volume of the cylinder (which is constant for the co-moving observers). In this calculation we have made use of the relation $dv /d\tau = \mathtt{a}/\gamma^2$ for an observer at the bottom of the cylinder, who is experiencing constant proper acceleration $\mathtt{a}$.

Naively, we ought to be able to arrive at the same result by considering the net Poynting flux crossing the boundary of the cylinder. Before we calculate this, it is interesting to understand the mechanism by which the Poynting vector carries energy into the cylinder. As noted earlier, on any given co-moving simultaneity the relative velocity, $v$, is constant along the cylinder and, thus, the proper Poynting vector is the same at the top and bottom of the cylinder. Since no flux leaves or enters the sides of the cylinder, this suggests that the net Poynting flux is zero. Recall, however, that because of the acceleration, proper time advances more quickly at the top of the cylinder relative to the bottom. This results in a greater proper time-integrated flux entering the top of the cylinder compared to that leaving the bottom of the cylinder---in other words, a net accumulation of electromagnetic energy. Net electromagnetic energy enters the cylinder because of the time dilation effect associated with the acceleration. With the magnitude of the Poynting vector given above, we find
\begin{eqnarray} \label{E(2)}
\Delta E_{\rm Poynting} = |\vec{S}| A \left( \Delta \tau^\prime - \Delta \tau \right)= 2 \beta\gamma^2 \left( \frac{E^2}{8\pi } V\right) \frac{\mathtt{a}\Delta\tau}{c}.
\end{eqnarray}
Observe that the net Poynting flux, equation (\ref{E(2)}), accounts for only {\it half} of the change in electromagnetic energy inside the cylinder, equation (\ref{E(1)}). Apparently, we have a paradox.

\subsection*{Paradox Resolution}\label{Paradox}

To understand the missing piece of this puzzle, let us first consider a simple problem in classical mechanics.  Imagine an object with rest mass $m$ from the point of view of a reference frame moving towards it with instantaneous speed $v$. In this reference frame, we would consider the object to have total energy $E =\sqrt{ m^2 c^4 + c^2 p^2}$, where $p = \gamma m v$ is the object's instantaneous momentum relative to the frame. Now, if our frame is accelerating at a proper rate $\mathtt{a}$, it is easy to show that, in a proper time $\Delta \tau$, the object's total energy changes by an amount $\Delta E = \mathtt{a} p \Delta \tau$. Thus, the total energy of the object increases solely as a result of the acceleration of our frame relative to the object.

The resolution to the paradox follows the same line of reasoning: as the cylinder accelerates relative to the existing momentum inside, there is an increase in energy proportional to this momentum times the acceleration. The momentum in this case belongs to the electromagnetic field contained in the cylinder, and is proportional to the Poynting vector:
\begin{eqnarray}
p =  \frac{1}{c^2} |\vec{S}| V = \beta \gamma^2 \left(\frac{E^2}{4\pi c } V\right).
\end{eqnarray}
(As a consistency check, observe that as the electromagnetic energy, $e = uV$, inside the cylinder increases, so does the electromagnetic momentum, $p$, such that $c^2$ times the invariant mass,
\begin{eqnarray}
m c^2 = \sqrt{e^2 - c^2 p^2} = \frac{E^2}{8\pi} V,
\end{eqnarray}
remains constant, as expected, and equal to the electromagnetic energy inside the cylinder when $v=0$.)

Now, for the problem at hand, the contribution to the energy change due to this effect is just
\begin{eqnarray}  \label{E(3)}
\Delta E_{\rm Bulk}  = \mathtt{a} p\, \Delta \tau = 2 \beta\gamma^2 \left(\frac{E^2}{8\pi } V\right)    \frac{\mathtt{a}\Delta\tau}{c}.
\end{eqnarray}
Notice that this energy change combined with the accumulation of Poynting flux due to the acceleration-induced time dilation along the length of the cylinder, $\Delta E_{\rm Poynting}$, now precisely matches that calculated from the energy density, i.e., $\Delta E_{\rm Poynting} + \Delta E_{\rm Bulk} = \Delta E$. Thus we have found the missing piece of the puzzle in the apparent paradox. It is not sufficient to consider the change in energy inside the cylinder based solely on the accumulation of Poynting flux across the surface; one must also take into account a ``momentum times acceleration" term due to the frame accelerating relative to an existing momentum. The inclusion of this {\it bulk} term then resolves the apparent paradox but comes with a cost---it does away with the usual {\it ``change in energy equals net energy flux through the boundary"} picture of a conservation law.

\section{Local Conservation Laws}\label{Local}

Let us now look more generally at the notion of conservation laws in the context of both special and general relativity.  Consider a smooth four-dimensional manifold, $\mathcal{M}$, endowed with a Lorentzian spacetime metric, $g_{ab}$, and associated covariant derivative operator, $\nabla_a$. In the presence of a non-zero matter stress-energy-momentum tensor, $T^{ab}$, we can construct the identity
\begin{equation}\label{BulkDifferential}
\nabla_a ( T^{ab}\Psi_b ) = ( {\nabla_a T^{ab}} ) \Psi_b + T^{ab} \nabla_{(a} \Psi_{b)}.
\end{equation}
This identity gives a differential conservation law for the current $Q^a = - T^{ab} \Psi_b$, whose physical interpretation depends on the choice of the weighting vector, $\Psi^a$. Integrating both sides of this identity over a finite four-volume, $\mathcal{V}$, we have
\begin{eqnarray} \label{BulkIntegrated}
\frac{1}{c} \int\limits_{\Sigma_f - \Sigma_i}  d{\Sigma}\,  T^{ab} u^\Sigma_a \Psi_b = \int\limits_{\mathcal{B}}  d \mathcal{B} \, T^{ab} n_a \Psi_b - \int\limits_{{\mathcal V}}  d{\mathcal V}\, \left[ \left( \nabla_a T^{ab} \right) \Psi_b + T^{ab} \left( \nabla_{(a} \Psi_{b)} \right)  \right].
\end{eqnarray}
On the left hand side, $\Sigma_i$ and $\Sigma_f$ are the initial and final three-dimensional spatial volume ``end caps" of $\mathcal{V}$, with timelike future-directed unit normal vector $\frac{1}{c}u^a_\Sigma$. On the right hand side, $\mathcal B$ is a three-dimensional timelike worldtube spanning the boundaries of the end caps, $\partial \Sigma_i$ and $\partial \Sigma_f$, with spacelike outward-directed unit normal vector, $n^{a}$, and induced Lorentzian three-metric $\gamma_{ab}=g_{ab}-n_{a}n_{b}$.

In general, the left hand side of equation~(\ref{BulkIntegrated}) gives the change in a quantity contained inside a spatial volume (e.g., electromagnetic energy) over some time interval, while the first term on the right hand side corresponds to matter fluxes across the boundary of that spatial volume during that time interval (e.g., electromagnetic Poynting flux). This is the desired form of a conservation law. However, there are two additional terms on the right hand side that are four-dimensional bulk integrals. The first of these involves $\nabla_a T^{ab}$, which is zero in general relativity; however in special relativity it will in general not vanish when $T^{ab}$ does not include all of the matter fields (e.g., when it does not include the electric four-current source of an electromagnetic field). The second term is present when $\Psi^a$ is not a Killing vector and, as it turns out, is crucial in resolving the apparent paradox above. To gain a better understanding of the terms in equation~(\ref{BulkIntegrated}) let us look at an example.

\subsection*{Example: Electromagnetism}

Consider a general electromagnetic field, which can be decomposed as
\begin{eqnarray}
F^{ab} = \frac{2}{c} u^{[a} E^{b]} + \epsilon^{ab}_{\phantom{ab}c} B^c,
\end{eqnarray}
where $u^a$ is the four-velocity of a volume-filling, three-parameter family of observers who see proper electric and magnetic fields $E^a$ and $B^a$, respectively, and $\epsilon_{bcd}=\frac{1}{c}u^{a}\epsilon_{abcd}$ is the spatial three-volume element orthogonal to the observers' worldlines. Denoting the spatial three-metric orthogonal to the observers' worldlines as $h_{ab} = g_{ab} + \frac{1}{c^2} u_a u_b$, the electromagnetic stress-energy-momentum tensor can be decomposed as
\begin{eqnarray}\label{EMStressTensor}
T^{ab} = \frac{1}{4\pi} \left\{ \frac{1}{2 c^2} u^a u^b \left( E^2 + B^2 \right) + \frac{2}{c} u^{(a} \epsilon^{b)}_{\phantom{b)}cd} E^c B^d - \left[ E^a E^b +B^a B^b -\frac{1}{2} h^{ab} \left( E^2 + B^2 \right) \right] \right\}
\end{eqnarray}
Also, it follows from Maxwell's equations that
\begin{eqnarray}
\nabla_a T^{ab} = j_a F^{ab},
\end{eqnarray}
where $j^a$ is the electric four-current source of the electromagnetic field.

For simplicity we will work with a congruence that has zero twist, i.e., one for which the four-velocity $u^a$ is hypersurface orthogonal, and assume that $u^a = u_\Sigma^a$ on $\Sigma_i$ and $\Sigma_f$. This reduction in generality does not affect our conclusions---it just makes the formulas simpler and more transparent. Choosing as our weighting vector $\Psi^a = \frac{1}{c} u^a$, equation (\ref{BulkIntegrated}) becomes an energy conservation equation, the individual terms of which are
\begin{align}
\frac{1}{c} T^{ab} u^\Sigma_a \Psi_b &= u, \label{Energy}\\
T^{ab} n_a \Psi_b &= - \frac{1}{c} \, n_a S^a, \\
\left( \nabla_a T^{ab} \right) \Psi_b &=  j_a E^a, \\
T^{ab} \left( \nabla_{(a} \Psi_{b)} \right) &= \frac{1}{c} \,  a_a \mathbb{P}^a - \mathbb{T}^{ab} K_{ab}.  \label{BulkTransfer2}
\end{align}
Here $u = \frac{1}{8\pi} \left( E^2 + B^2 \right)$ and $S^a = \frac{c}{4\pi} \epsilon^a_{\phantom{a}bc} E^b B^c$ are the proper electromagnetic energy density and Poynting vector, respectively, and $j^a$ is the electric four-current vector, as before. $\mathbb{P}^a = \frac{1}{c^2}  S^a$ is the proper electromagnetic momentum density, and $\mathbb{T}^{ab} = - h^a_{\phantom{a}c} h^b_{\phantom{b}d} T^{cd}$ is the spatial, three-dimensional Maxwell stress tensor. $a^a = u^b \nabla_b u^a$ is the observers' four-acceleration, and $K_{ab} = \frac{1}{c} h_{(a}^{\phantom{(a}c} h_{b)}^{\phantom{b)}d} \nabla_c u_d$ is the observers' spatial, three-dimensional strain rate tensor, measuring the expansion and shear of their congruence. Inserting equations (\ref{Energy}-\ref{BulkTransfer2}) into equation (\ref{BulkIntegrated}) gives a general relativistic version of Poynting's theorem for a hypersurface orthogonal congruence of observers:
\begin{eqnarray} \label{EMBulkIntegrated}
\int\limits_{\Sigma_f - \Sigma_i}  d{\Sigma}\,  u = - \frac{1}{c} \int\limits_{\mathcal{B}}  d \mathcal{B} \,   n_a S^a -  \int\limits_{{\mathcal V}}  d{\mathcal V}\, \left[  j_a E^a +  \frac{1}{c} \,  a_a \mathbb{P}^a - \mathbb{T}^{ab} K_{ab}    \right].
\end{eqnarray}
Notice that the standard form of Poynting's theorem is recovered when $u^a$ is a timelike Killing vector, in which case the last two bulk terms on the right hand side, coming from equation (\ref{BulkTransfer2}), vanish. (The bulk term involving $j_a E^a$ remains, however, and represents an energy transfer between the electromagnetic field and its electric four-current source.)  When $u^a$ is {\it not} a Killing vector, however, these two bulk terms provide additional mechanisms for energy transfer. The $a_a \mathbb{P}^a$ term represents change in energy due to the frame accelerating relative to an existing electromagnetic momentum in the system (as was seen in resolving the apparent paradox of \S\ref{Paradox}). The $\mathbb{T}^{ab} K_{ab}$ term represents the Maxwell stress, $\mathbb{T}^{ab}$, doing work against the strain, $K_{ab}$, of the three-parameter congruence. Since we would like to generalize beyond the case where $u^a$ is a Killing vector, both of these energy transfer mechanisms will be important.

\subsection*{Apparent Paradox Revisited}

We will now illustrate the use of the local conservation law, equation (\ref{BulkIntegrated}), to resolve the apparent paradox discussed in \S\ref{Paradox}, generalized slightly to allow for a time-dependent acceleration along the $Z$-axis.\footnote{The analysis can be generalized to arbitrary acceleration; the limit to acceleration along the $Z$-axis is chosen purely for notational simplicity.} Moreover, to facilitate comparison with the general relativistic calculation in the next section (on {\it quasi}local conservation laws), we will switch from a right circular cylinder of length $L$ and cross-sectional area $A$ to a round sphere of areal radius $r$. The reason for this switch is that the general relativistic calculation involves the extrinsic curvature of the boundary of our spatial volume, and the sharp corners of a cylinder introduce an unnecessary technical complication.

Let us denote the Minkowski coordinates of an inertial reference frame in flat spacetime by $X^a = (X^0, X^I)=(cT, X,Y,Z)$, $I=1,2,3$. Let the embedding $X^a=\xi^a(\tau)$ define an accelerated, timelike worldline, ${\cal C}_0$, where $\tau$ is proper time, and let $e_0^{\phantom{0}a}(\tau)=\frac{1}{c}d\xi^a(\tau)/d\tau$ denote the unit vector tangent to ${\cal C}_0$. We will construct a three-parameter family of accelerated observers in the neighborhood of this worldline using the coordinate transformation
\begin{eqnarray} \label{Worldline}
X^a (x) = \xi^a (\tau) + r r^I (\theta, \phi) e_I^{\phantom{I}a}(\tau).
\end{eqnarray}
Here $x^\alpha = (x^0,x^1,x^i)=(\tau, r, \theta, \phi)$ are coordinates adapted to the congruence of observers: the spherical coordinates $(r, \theta, \phi)$ are parameters that label the observers' worldlines, and $\tau$ is a time parameter along the worldlines which in general is {\it not} proper time, except for the observer at $r=0$ (i.e., there is a nontrivial lapse function, $N$, which will be given shortly). Also, $r^I (\theta, \phi) = (\sin{\theta} \cos{\phi}, \sin{\theta} \sin{\phi}, \cos{\theta})$ are the usual direction cosines in a spherical coordinate system, and $e_I^{\phantom{I}a}(\tau)$ is a Fermi-Walker transported spatial triad defined on ${\cal C}_0$ that is orthogonal to $e_0^{\phantom{0}a}(\tau)$.

Specializing to the case relevant to our apparent paradox, we let ${\cal C}_0$ (i.e., the observer at $r=0$) undergo proper acceleration $\mathtt{a}(\tau)$ along the $X^3$ (i.e., $Z$) axis, in which case our tetrad has the form:
\begin{align}
e_0^{\phantom{0}a}(\tau) &= \cosh{\alpha (\tau)} \, \delta_0^a + \sinh{\alpha (\tau)} \, \delta_3^a, \label{tetrad_0}\\
e_1^{\phantom{1}a}(\tau) &= \delta_1^a, \label{tetrad_1}\\
e_2^{\phantom{2}a}(\tau) &= \delta_2^a, \label{tetrad_2}\\
e_3^{\phantom{3}a}(\tau) &= \sinh{\alpha (\tau)} \, \delta_0^a + \cosh{\alpha (\tau)} \, \delta_3^a, \label{tetrad_3}
\end{align}
where $\alpha (\tau) = \frac{1}{c}\int^\tau \, \mathtt{a}(t) \,dt$. A short calculation reveals that the metric in the observer-adapted coordinate system is
\begin{equation}\label{AdaptedCoordMetric}
g_{\alpha\beta}=
\left(
\begin{array}{ccc}
-c^2 N^2 & 0 & 0 \\
0 & 1 & 0 \\
0 & 0 & r^2\mathbb{S}_{ij}
\end{array}
\right),
\end{equation}
where $\mathbb{S}_{ij}={\rm diagonal}(1,\sin^2\theta)$ is the metric on the unit round sphere, and $N(x) = 1 + \frac{1}{c^2}\mathtt{a}(\tau) r \cos \theta$ is the lapse function. The observers' four-velocity is defined as $u^a(x)=N^{-1}(x)(\partial/\partial \tau)^a=c e_0^{\phantom{0}a}(\tau)$, with resulting four-acceleration $a^a(x)=N^{-1}(x)\mathtt{a}(\tau)e_3^{\phantom{3}a}(\tau)$. Comparing with the discussion of the paradox given in {\S}\ref{Paradox}, in the context of an accelerating cylinder, the nontrivial lapse function, $N(x)$, is the analogue of our earlier relation $\Delta\tau^\prime =(1+\mathtt{a}L/c^2)\Delta\tau$, i.e., when $\mathtt{a}(\tau)>0$, proper time passes more quickly for observers above the plane $\theta=\pi/2$ relative to those below it. Also, the magnitude of the observers' proper acceleration, $N^{-1}(x)\mathtt{a}(\tau)$, is the analogue of our earlier relation $\mathtt{a}^\prime=\mathtt{a}/(1+\mathtt{a}L/c^2)$, i.e., when $\mathtt{a}(\tau)>0$, observers above the plane $\theta=\pi/2$ experience less proper acceleration compared to those below it. Finally, by inspection of the metric in equation (\ref{AdaptedCoordMetric}), it is clear that the observers are accelerating {\it rigidly}, in the sense that the orthogonal distance between all nearest neighbor pairs of observers remains constant in time, as in the Bell spaceship paradox (despite the time-dependent acceleration). Thus, the three-dimensional strain rate tensor, $K_{ab}$ in equation~(\ref{BulkTransfer2}), is zero. (This is a very special case. In general, it is not possible to find a volume-filling congruence of observers for which $K_{ab}=0$, a point we will return to in the next section.)

As in {\S}\ref{Paradox}, we now introduce a constant, uniform electric field of magnitude $E$ in the positive $X$-direction, perpendicular to the observers' motion: $E^a = E \delta_1^a$ (and $B^a$ and $j^a$ vanish, at least in the region of the observers' congruence). It follows from equation (\ref{EMStressTensor}) that the electromagnetic stress-energy-momentum tensor is $T^{ab}=(E^2/8\pi)\times{\rm diagonal}(1,-1,1,1)$. Choosing $\Psi^a = \frac{1}{c} u^a$ in equation (\ref{BulkIntegrated}) makes this an energy conservation equation. With $u_{\Sigma}^a=u^a$ and $n^a=r^I e_I^{\phantom{I}a}$ we find that the individual terms in equation (\ref{BulkIntegrated}) are
\begin{align}
\frac{1}{c} T^{ab} u^\Sigma_a \Psi_b &= \left( 1 + \beta^2 \right) \gamma^2 \frac{E^2}{8\pi}, \\
T^{ab} n_a \Psi_b &= \beta \gamma^2 \frac{E^2}{4\pi} \cos \theta, \\
\left( \nabla_a T^{ab} \right) \Psi_b &= 0, \\
T^{ab} \left( \nabla_{(a} \Psi_{b)} \right) &= -\beta \gamma^2 \frac{E^2}{4\pi} \frac{\mathtt{a}(\tau)}{c^2}\frac{1}{N}  , \label{BulkEM2}
\end{align}
where we have introduced the usual Lorentz transformation parameters: $\gamma(\tau) = \cosh \alpha(\tau)$ and $\beta(\tau) = \tanh \alpha(\tau)$ such that $\gamma = 1/\sqrt{1 - \beta^2}$. Observe that the right hand side of equation (\ref{BulkEM2}) is due entirely to the $a_a \mathbb{P}^a$ term in equation (\ref{BulkTransfer2}) since, as noted above, the collective rigidity of the observers' motion means that the strain rate tensor, $K_{ab}$, vanishes in this case. In other words, this last contribution is entirely the result of a change in energy due to the frame accelerating with respect to the existing electromagnetic momentum in the system. Putting these results together, the conservation law reads
\begin{equation} \label{ParadoxConservationLaw}
\int\limits_{\Sigma_f - \Sigma_i}  dr \, r^2 \, d{\mathbb{S}}\,   \left( 1 + \beta^2 \right) \gamma^2 \frac{E^2}{8\pi} = \int\limits_{\mathcal{B}}  c\,d\tau\,r^2\,d\mathbb{S} \,  \beta \gamma^2 \frac{E^2}{4\pi} \cos \theta \left(1 + \frac{1}{c^2}\mathtt{a}(\tau) r \cos \theta\right)  + \int\limits_{{\mathcal V}}  c\,d\tau\,dr\,r^2\,d\mathbb{S}\, \beta \gamma^2 \frac{E^2}{4\pi} \frac{\mathtt{a}(\tau)}{c^2},
\end{equation}
where the lapse function in the bulk term, $T^{ab} \left( \nabla_{(a} \Psi_{b)} \right)$---see equation~(\ref{BulkEM2}), has been canceled by that in the volume element, $d\mathcal{V} = c\,d\tau\,dr\,r^2\,d\mathbb{S}\,N$.  Note also that we have used $d\Sigma =  dr \, r^2 \,d\mathbb{S}$  and $ d\mathcal{B} = c\,d\tau\,r^2\,d\mathbb{S} \,N $, where $d\mathbb{S} = \sin \theta \,d\theta\, d\phi$ is the area element on the unit round sphere.

In order to compare directly with the results in {\S}\ref{Paradox}, let the time interval be infinitesimal: $\tau_f-\tau_i=\Delta\tau$. Using the relation $\frac{d}{d\tau} \left( (1 + \beta^2) \gamma^2  \right) = 4 \beta \gamma^2\frac{\mathtt{a}}{c} $, the left hand side of equation~(\ref{ParadoxConservationLaw}) then integrates to precisely the same change in electromagnetic energy, $\Delta E$, given in equation~(\ref{E(1)}), except with the cylinder volume, $V=AL$, replaced with the sphere volume, $V=\frac{4}{3} \pi r^3$. Also, the $\Delta\tau$ in the equation now refers to the proper time elapsed for the observer at the center of the sphere, instead of an observer at the bottom of the cylinder. The Poynting flux integral over $\mathcal{B}$ on the right hand side of equation~(\ref{ParadoxConservationLaw}) has two terms. The first term (proportional to $\cos\theta$) integrates to zero over the angles, which is analogous to the proper Poynting vector in {\S}\ref{Paradox} being the same at the top and bottom of the cylinder such that, at lowest order, flux in equals flux out. At the next order in $r$, however, the nontrivial lapse function is responsible for a $\cos^2\theta$ term that does {\it not} integrate to zero. As in {\S}\ref{Paradox}, it is this non-isotropic time dilation that allows for a nonzero accumulation of Poynting flux. Evaluating this integral we find precisely the same change in electromagnetic energy, $\Delta E_{\rm Poynting}$, given in equation~(\ref{E(2)}), except with $V$ and $\Delta\tau$ reinterpreted as discussed above. Finally, the bulk integral over $\mathcal{V}$ on the right hand side of equation~(\ref{ParadoxConservationLaw}), which measures the change in energy due to the acceleration of the frame relative to the existing electromagnetic momentum in the system, evaluates to precisely the same change in electromagnetic energy, $\Delta E_{\rm Bulk}$, given in equation~(\ref{E(3)}), except with $V$ and $\Delta\tau$ again reinterpreted as discussed above.

This example illustrates that any local conservation law constructed from the matter stress-energy-momentum tensor, $T^{ab}$, as in equations~(\ref{BulkDifferential}) and (\ref{BulkIntegrated}), in general contains {\it two} terms responsible for the change in a physical quantity inside a volume of space: the first is a three-dimensional surface flux integral over $\cal B$, as one might expect, and the second is a four-dimensional bulk integral over $\cal V$. It is the addition of this bulk integral, which naively we did not expect, that resolves the apparent paradox introduced in \S\ref{Paradox}. But is such a bulk integral really necessary? In the next section we will see how to naturally convert this bulk integral into a surface flux integral, and at the same time generalize the conservation law in equations~(\ref{BulkDifferential}) and (\ref{BulkIntegrated}) to include gravitational effects.

\section{Quasilocal Conservation Laws}\label{Quasilocal}

We will now argue that the local conservation law given in equations~(\ref{BulkDifferential}) and (\ref{BulkIntegrated}) is defective in two respects. First, even when matter is locally covariantly conserved, i.e., $\nabla_a T^{ab}=0$ (which is always true in general relativity), the local conservation law contains a nontrivial bulk term when $T^{ab}$ is present and $\Psi^a$ is not a Killing vector. Insofar as a generic spacetime does not admit any Killing vectors, when $T^{ab}$ is present this bulk term is generically present. But this bulk term violates what we usually think of as a conservation law, i.e., that the change in some physical quantity over a period of time is equal to some related surface flux integral over that period of time. Is there a natural and general way to convert this bulk term into a surface flux term?

To help motivate this question, let us return for a moment to the ``momentum times acceleration" bulk term that resolves the apparent paradox introduced in \S\ref{Paradox}. We imagine being inside an accelerating box in flat spacetime that contains a freely-floating, massive object that appears to be accelerating toward us; the object's kinetic energy (relative to us) increases due to the acceleration of our frame. We ask: Where does the increasing kinetic energy come from? This might sound like a silly question---after all, energy is frame-dependent, and we are just changing the frame! However, the question is perhaps not so silly when we ask it in the context of the equivalence principle. Instead of being inside an accelerating box, we could imagine that the box is at rest in a uniform gravitational field, and that the object is experiencing an acceleration toward us due to the ``force" of gravity. This ``force" acting through a distance represents an energy transfer mechanism from the gravitational field energy to the kinetic energy of the object. So it might be possible to convert the ``momentum times acceleration" bulk term into some kind of surface flux term representing {\it gravitational} energy entering the box from the outside. In the context of general relativity, i.e., when we properly account for the frame dragging produced by the object in motion, we will see that this is exactly what happens.

The second defect, which is related to the first, is that the local conservation law given in equations~(\ref{BulkDifferential}) and (\ref{BulkIntegrated}) cannot properly account for gravitational effects since it is {\it homogeneous} in $T^{ab}$. In any vacuum spacetime region where $T^{ab}=0$, the local conservation law has nothing to say. For example, we can imagine a vacuum region of space containing gravitational energy that is flowing in or out, to which the local conservation law is completely blind. This is, of course, to be expected, given that gravitational energy is not localizable. It seems that we need a term {\it like} $\nabla_{(a}\Psi_{b)}$, which can be thought of as a measure of the presence of interesting gravitational physics, e.g., a non-stationary spacetime, and that this term should be coupled not to $T^{ab}$, but rather some kind of {\it quasi}local stress-energy-momentum tensor that represents both matter {\it and} gravity, so that it can be nonzero even when $T^{ab}=0$. We will presently construct a general quasilocal conservation law with precisely these properties, that will remove this second defect and, by its very construction, will also automatically remove the first defect, in an interesting and subtle way, exactly as anticipated in the previous paragraph.

Let us consider an identity exactly analogous to equation (\ref{BulkDifferential}), except defined in the {\it three}-dimensional spacetime of the timelike worldtube, $\mathcal B$, i.e., the history of the boundary of a three-dimensional system. Then, for an arbitrary vector field, $\psi^a$, tangent to $\mathcal B$, we have the identity
\begin{equation}\label{QuasilocalDifferential}
D_a ( T_{\mathcal B}^{ab}\psi_b ) = ( D_a T_{\mathcal B}^{ab} ) \psi_b + T_{\mathcal B}^{ab} D_{(a} \psi_{b)},
\end{equation}
where $D_a$  is the covariant derivative with respect to the three-metric, $\gamma_{ab} = g_{ab} - n_a n_b $, induced in $\mathcal B$. In place of the four-dimensional {\it matter} stress-energy-momentum tensor, $T^{ab}$, used in equation (\ref{BulkDifferential}), we have inserted the three-dimensional {\it total} (matter plus gravitational) stress-energy-momentum tensor, $T_{\mathcal B}^{ab}$, defined by Brown and York \cite{BY1993}. This {\it quasi}local stress-energy-momentum tensor is defined as $T_{\mathcal B}^{ab} = -\frac{1}{\kappa}\,\Pi^{ab}$, where $\Pi^{ab}$ is the gravitational momentum canonically conjugate to the three-metric $\gamma_{ab}$ on $\mathcal B$, and $\kappa = 8 \pi G / c^4$. $\Pi^{ab}$, in turn, is equal to $\Theta_{ab} - \Theta \gamma_{ab}$, where $\Theta_{ab}=\gamma_a^{\phantom{a}c}\nabla_c n_b$ is the extrinsic curvature of $\mathcal B$. It will be useful to decompose the Brown-York quasilocal stress-energy-momentum tensor into energy, momentum, and stress surface densities, respectively, as:
\begin{align}\label{SurfaceSEMcomponents}
{\mathcal E} & = \frac{1}{c^2} u^a u^b T^{\mathcal B}_{ab} & {\rm [Energy/Area]},\\
{\mathcal P}_a & = - \frac{1}{c^2} \sigma_{a}^{\phantom{a}b}u^{c}T^{\mathcal B}_{bc} & {\rm [Momentum/Area]},\\
{\mathcal S}_{ab} & = \sigma_{a}^{\phantom{a}c}\sigma_{b}^{\phantom{b}d}T^{\mathcal B}_{cd} & {\rm [Force/Length]},
\end{align}
where $u^a$ is the four-velocity of a two-parameter family of observers residing at the boundary of a spatial volume; i.e., the worldtube boundary, $\mathcal B$, is the congruence of the integral curves of $u^a$. The spatial two-metric, $\sigma_{ab} = g_{ab} - n_a n_b + \frac{1}{c^2} u_a u_b$, projects tensors into the space orthogonal to both $u^a$ and $n^a$, i.e., tangent to the spatial two-surface the observers reside on. Integrating equation (\ref{QuasilocalDifferential}) over a section of the worldtube, $\mathcal{B}$, bounded by initial and final spacelike slices, $\mathcal{S}_i$ and $\mathcal{S}_f$, we have
\begin{equation} \label{QuasilocalIntegrated}
\frac{1}{c} \int\limits_{\mathcal{S}_f - \mathcal{S}_i}  d{\mathcal{S}}\,  T_{\mathcal B}^{ab} u^{\mathcal{S}}_a \psi_b = \int\limits_{\mathcal{B}}  d \mathcal{B} \, \left[  T^{ab} n_a \psi_b - T_{\mathcal B}^{ab} \left( D_{(a} \psi_{b)} \right)  \right].
\end{equation}
On the left hand side, $d{\mathcal{S}}$ is the surface area element on $\mathcal{S}_i$ and $\mathcal{S}_f$, and $\frac{1}{c}u_{\mathcal{S}}^a$ is the timelike future-directed unit vector normal to $\mathcal{S}_i$ and $\mathcal{S}_f$ (and tangent to $\mathcal B$).

It is interesting to compare this quasilocal conservation law to the local conservation law in equation (\ref{BulkIntegrated}). The left hand side of both equations gives the change in a physical quantity (e.g., energy) contained in a spatial three-volume. The first key difference is that the left hand side of equation (\ref{QuasilocalIntegrated}) is {\it quasilocal}: it is an integral of a {\it surface} density (e.g., energy per unit area) over the two-surface boundary of the three-volume. Unlike the local volume density, the quasilocal surface density has no meaning by itself; only the quasilocal surface density integrated over a closed two-surface is physically meaningful. The second key difference is that the integrated surface density includes contributions from both matter {\it and} gravity (e.g., gravitational energy) \cite{BY1993}.

The first term on the right hand side of equation (\ref{QuasilocalIntegrated})---the matter flux term---is identical to that in equation (\ref{BulkIntegrated}) (when $\Psi^a=\psi^a$ is tangent to $\mathcal B$), except its origin is very different. Unlike equation (\ref{BulkDifferential}), equation (\ref{QuasilocalDifferential}) is purely geometrical, involving only the intrinsic and extrinsic geometry of $\mathcal B$, with no reference to matter. The matter stress-energy-momentum tensor, $T^{ab}$, enters equation (\ref{QuasilocalIntegrated}) by first applying the Gauss-Codazzi identity ($D_a T_{\mathcal B}^{ab} = -\frac{1}{\kappa} D_a \Pi^{ab} = -\frac{1}{\kappa} n_a G^{ab}$, where $G_{ab}$ is the Einstein tensor), and then applying the Einstein equation ($G_{ab} = \kappa T_{ab}$). Thus, equation (\ref{QuasilocalIntegrated}) forbids us from ignoring the fact that matter distorts the spacetime geometry; for example, it is meaningless to analyze an electromagnetic field in {\it flat} spacetime, as we did in {\S}\ref{Local}. Notice also that, because the boundary of a boundary is zero, there is actually {\it no} term in the quasilocal conservation law that is analogous to the matter flux term in equation (\ref{BulkIntegrated}). The matter flux term in equation (\ref{QuasilocalIntegrated}) arises through an entirely different mechanism: it is actually a surface flux analogue of the $j_a E^a$ bulk term in Poynting's theorem. The latter term arises when $\nabla_a T^{ab}\not=0$, which is analogous to $D_a T_{\mathcal B}^{ab} \not=0$ in the quasilocal conservation law. Specifically, $\nabla_a T^{ab}=j_a F^{ab}$ represents local energy-momentum transfer from the electromagnetic field to its sources, whereas $D_a T_{\mathcal B}^{ab} = - n_a T^{ab}$ represents quasilocal energy-momentum transfer from the matter fields to the ``system" contained in $\mathcal B$.

The last term on the right hand side of the quasilocal conservation law in equation (\ref{QuasilocalIntegrated}) is analogous to the last term in equation (\ref{BulkIntegrated}). To make the comparison more detailed, let us look at energy conservation by taking $\psi^a = \frac{1}{c} u^a$. Using the decomposition in equation~(\ref{SurfaceSEMcomponents}) we find
\begin{equation}\label{QuasilocalTransfer}
\frac{1}{c}\,T_{\mathcal B}^{ab} \left( D_{(a} u_{b)} \right) = \frac{1}{c}\,\alpha_a \mathcal{P}^a +  \mathcal{S}^{ab} \theta_{ab},
\end{equation}
where $\alpha^a = \sigma^a_{\phantom{a}b} a^b$ is the projection of the observers' four-acceleration tangent to $\mathcal B$, and $\theta_{ab} = \frac{1}{c} \sigma_{(a}^{\phantom{(a}c} \sigma_{b)}^{\phantom{b)}d} \nabla_{c} u_{d}$ is the strain rate tensor describing the time development (i.e., expansion and shear) of their congruence. The structure of equation~(\ref{QuasilocalTransfer}) is identical to that of equation~(\ref{BulkTransfer2}), except it is one dimension lower: surface, rather than volume, i.e., quasilocal, rather than local. By the very construction of the quasilocal conservation law, the volume densities $a_a \mathbb{P}^a$ and $- \mathbb{T}^{ab} K_{ab}$ in equation~(\ref{BulkTransfer2}) automatically become surface densities, $\alpha_a \mathcal{P}^a$ and $\mathcal{S}^{ab} \theta_{ab}$, respectively, i.e., surface fluxes. This removes the first defect of local conservation laws discussed at the beginning of this section.

Let us first focus on the quasilocal ``momentum times acceleration" term, $\alpha_a \mathcal{P}^a$. Unlike the momentum volume density, $\mathbb{P}^a$, the momentum surface density, $\mathcal{P}^a$, is purely geometrical: it is a component of the extrinsic curvature of $\mathcal B$ closely related to a vector, $\omega^a$, that measures the precession rate of the observers' local ``radial" vector, $n^a$, relative to inertial gyroscopes. As shown in more detail in reference \cite{EMM2011}, the precise relation is: $\mathcal{P}^a = \frac{c^2}{8\pi G}\,\epsilon^{a}_{\phantom{a}b}\omega^{b}$, where $\epsilon_{ab} = \frac{1}{c}\epsilon_{abcd}u^c n^d$ is the two-dimensional spatial volume form orthogonal to both $u^a$ and $n^a$. This allows us to write:
\begin{equation}\label{GravitationalPoynting}
\alpha_a \mathcal{P}^a = \frac{c^2}{8\pi G}\epsilon_{ab}\alpha^a \omega^b,
\end{equation}
which is reminiscent of the normal component of the electromagnetic Poynting vector, $\hat{n}\cdot \frac{c}{4\pi}\vec{E}\times\vec{B}$. In fact, as we argue in more detail in reference \cite{EMM2011}, we believe that $\alpha_a \mathcal{P}^a$ is the exact, purely geometrical, operational expression for gravitational energy flux. So the ``momentum times acceleration" bulk term, $a_a \mathbb{P}^a$, has really become a gravitational energy surface flux term, exactly as anticipated in the equivalence principle argument given near the beginning of this section. In the context of general relativity, the presence of momentum (matter or gravitational) flowing through the system causes the observers' local ``radial" vector to precess relative to inertial gyroscopes (i.e., a frame dragging effect), and the vector cross product between this precession rate (the gravitational analogue of a magnetic field) and the observers' acceleration (the gravitational analogue of an electric field) corresponds to a flow of gravitational energy into the system.

Of course this frame dragging mechanism of energy transfer would be difficult to confirm experimentally because the frame dragging precession rate is smaller than the momentum density by the factor $\frac{c^2}{8\pi G}$. Nevertheless, the existence of this mechanism is not surprising given that it bears a close resemblance to a similar mechanism in electrodynamics. In the context of special relativity, consider a charged particle accelerating in a uniform electric field (for simplicity we will ignore radiation reaction, which doesn't affect the point of our discussion). We ask: Where does the particle's increasing kinetic energy come from? The answer is obviously ``the field." But what is the mechanism, exactly? Consider that the velocity field part of the magnetic field due to the particle's motion is localized near the particle, and circulates around its axis of motion. A moment's thought shows that the vector cross product between the external electric field (causing the particle's acceleration) and this magnetic velocity field (moving with the particle) points toward the axis of motion, and is maximum in the plane containing the particle. In other words, there is a Poynting vector representing a flow of energy from the field to the particle. Intuitively, one might guess that this answers our question. Surprisingly, it seems that a detailed analysis of this basic energy transfer mechanism in classical electrodynamics has been done only very recently \cite{Gaidukov}. The authors of this reference verify that our intuition is correct, at least in the ultra-relativistic limit.\footnote{In the slow-motion limit, only $2/3$ of the particle's kinetic energy is supplied by the Poynting flux; $1/3$ is supplied by a peculiar ``interference" effect between the external electric field and the velocity field part of the particle's electric field \cite{Gaidukov}. This subtlety is interesting, but it does not affect the spirit of our discussion.}

To the extent that the linearized approximation to general relativity is very similar to electrodynamics, one might expect essentially the same energy transfer mechanism to occur in gravitational physics, when, e.g., we replace the massive, charged particle accelerating in a uniform electric field with a massive (charged or uncharged) particle accelerating in a uniform gravitational field (or, via the equivalence principle, a frame accelerating toward a freely-floating particle). In the context of the linearized, slow motion approximation to general relativity, sometimes called {\it gravitoelectromagnetism} (for a review, see \cite{Mashhoon}), precisely such a mechanism has been confirmed; see, e.g., \cite{Krumm,Matos}. What's new in this paper is that this frame dragging energy transfer mechanism is confirmed in the full, nonlinear general theory of relativity with arbitrary matter and with no approximations; the gravitational energy flux, $\alpha_a \mathcal{P}^a$, is an exact, purely geometrical contribution to the energy flux that must be considered to properly explain a wide variety of phenomena, including phenomena as basic as a falling apple. We will compute an explicit example of this energy transfer mechanism in the next section.

Let us now focus on the quasilocal ``stress times strain" term, $\mathcal{S}^{ab} \theta_{ab}$, in equation~(\ref{QuasilocalTransfer}). The boundary observers' strain rate tensor, $\theta_{ab}$, is the two-dimensional analogue of the three-dimensional strain rate tensor, $K_{ab}$, in equation~(\ref{BulkTransfer2}), for a volume-filling set of observers. In the latter case, it is obvious that such a {\it three}-parameter family of observers cannot, in general, move {\it rigidly}, i.e., move in such a way as to maintain constant radar-ranging distances between all nearest neighbor pairs of observers. This is because the condition $K_{ab} = 0$ represents six differential constraints on three functions (the three independent components of the observers' four-velocity, $u^a$). In other words, the local conservation law generically contains a nontrivial bulk ``stress times strain" term, $- \mathbb{T}^{ab} K_{ab}$.

The same is {\it not} true for a {\it two}-parameter family of observers, because there the condition $\theta_{ab} = 0$ represents only {\it three} differential constraints on the same three functions. In reference \cite{EMM2011} we argue that, in a generic spacetime, it is always possible to find a two-parameter congruence of integral curves of $u^a$, comprising $\mathcal B$, that is expansion- and shear-free. Moreover, the degrees of freedom of motion left to such a congruence of boundary observers are precisely the same as those for rigid body motion in Newtonian space-time, viz., three translations and three rotations, each with arbitrary time dependence. In other words, in the quasilocal conservation law (but {\it not} in the local conservation law), it is always possible to choose a family of observers such that the ``stress times strain" term vanishes. We have called such a family of observers a {\it Rigid Quasilocal Frame}, or RQF for short. We refer the reader to references \cite{EMM2009, EMM2011} for a more detailed discussion of the properties of RQFs. For our present purposes we need only appreciate that, since a congruence of RQF observers is expansion- and shear-free, there are no extraneous fluxes entering or leaving the system due merely to a change in the size or shape of the system boundary. So all of the interesting physics lies in the $\alpha_a \mathcal{P}^a$ surface flux term discussed above. This is a key advantage to working with a quasilocal conservation law in general, and RQFs in particular. We will compute an explicit example of an RQF in the next section.

In summary, the quasilocal conservation law introduced in this section removes two defects of the familiar local conservation law: (1) The undesirable bulk terms in the local conservation law automatically appear as surface flux terms in the quasilocal conservation law, and (2) unlike the local conservation law, the quasilocal conservation law accounts for gravitational effects. For example, the quasilocal energy conservation law contains an energy transfer mechanism related to frame dragging that is needed to understand in detail what is actually happening with regards to energy flow when, say, an apple falls. In the next section we will apply the quasilocal energy conservation law to a general relativistic analogue of the apparent paradox introduced in {\S}\ref{Paradox}.

\subsection*{Quasilocal Approach to the Apparent Paradox}

To illustrate the explanatory power of the quasilocal approach let us now look at a general relativistic analogue of the apparent paradox introduced in \S\ref{Paradox}. We will work with a spacetime whose metric is $g_{ab}$ in coordinates $X^a = (X^0, X^I)=(cT, X,Y,Z)$, $I=1,2,3$, and which contains a covariantly constant (i.e., uniform) electric field of magnitude $E$ in the positive $X$-direction \cite{Bertotti,Stephani}:
\begin{equation}\label{ElectricSpacetime}
ds^2 = -\left(1 +  X^2/L^2 \right) c^2 dT^2 + \frac{dX^2}{\left( 1 +  X^2/L^2 \right)}+ \frac{dY^2}{\left( 1 - Y^2/L^2 \right)} + \left(1 -  Y^2/L^2 \right) dZ^2,
\end{equation}
where $L = c^2/\sqrt{G} E$ is the length scale of the geometry. It is important to note that the presence of the electric field is intimately connected to the spacetime curvature---we are no longer working in flat spacetime. This is readily verified by checking that there exist non-zero curvature invariants such as $R_{a b c d} R^{a b c d} = 8 /L^4$. Alternatively, it is easy to see that this spacetime is a product of two surfaces of constant curvature \cite{Bertotti,Stephani}.

Observe that $g_{ab}$ reduces to the Minkowski metric in the $X=0,Y=0$ plane. In this plane let us construct a timelike worldline, ${\mathcal C}_0$, parameterized by proper time, $\tau$, and defined by the embedding $X^a = \xi^a (\tau)$, that represents a fiducial observer undergoing proper acceleration $\mathtt{a}(\tau)$ along the $Z$-axis. As in \S\ref{Local}, along ${\mathcal C}_0$ we define a tetrad, $e_0^{\phantom{0}a}(\tau)$ and $e_I^{\phantom{I}a}(\tau)$, where $e_0^{\phantom{0}a}(\tau)=\frac{1}{c}\,d\xi^a(\tau)/\,d\tau$ denotes the unit vector tangent to ${\mathcal C}_0$, and the spatial triad, $e_I^{\phantom{I}a}(\tau)$, is Fermi-Walker transported along ${\mathcal C}_0$. Because $g_{ab}$ equals the Minkowski metric on $\mathcal{C}_0$, we can use the same coordinate components for this tetrad as given in equations~(\ref{tetrad_0})-(\ref{tetrad_3}). Around $\mathcal{C}_0$ we will construct a two-parameter family (two-sphere's worth) of accelerated observers, whose congruence of worldlines comprises a timelike worldtube, $\mathcal{B}$. Our notation will follow that used in \S\ref{Local}: The spacetime metric, $g_{ab}$, induces on $\mathcal{B}$ the spacelike outward-directed unit normal vector field, $n^a$, the Lorentzian three-metric, $\gamma_{ab} = g_{ab} - n_a n_b$, and the covariant derivative operator, $D_a$. Also, we let $u^a$ denote the observers' four-velocity, which is tangent to $\mathcal{B}$, and $\sigma_{ab} = g_{ab} - n_a n_b + \frac{1}{c^2} u_a u_b$ denote the observers' spatial two-metric.

In \S\ref{Local} we constructed a three-parameter (volume-filling) family of accelerated observers in flat spacetime in the neighborhood of ${\mathcal C}_0$ using the coordinate transformation given in equation~(\ref{Worldline}). For arbitrary fixed $r$, this coordinate transformation actually defines an exact RQF, i.e., a two-parameter family (two-sphere's worth) of observers who, despite their time-dependent acceleration, maintain constant radar-ranging distance from their nearest neighbors---they are moving {\it rigidly}. Indeed, inspection of equation~(\ref{AdaptedCoordMetric}) shows that $n_i$ and $u_i$ both vanish, so the observers reckon they are on a surface with constant spatial two-metric $\sigma_{ij} = g_{ij} = r^2\mathbb{S}_{ij}$, which is a round sphere of areal radius $r$. Now because spacetime is locally flat, at lowest order in $r$ we can start with the same coordinate transformation given in equation~(\ref{Worldline}), but because of the spacetime curvature, we must add corrections at higher order in $r$ to achieve the same RQF, i.e., to achieve $\sigma_{ij} = r^2\mathbb{S}_{ij}$. Thus we begin with the ansatz:
\begin{equation} \label{RQFWorldTube}
X^a (x) = \xi^a (\tau) + r r^I (\theta, \phi) e_I^{\phantom{I}a}(\tau) + \frac{r^3}{L^2}\left[
F^I(\tau,\theta,\phi) e_I^{\phantom{I}a}(\tau) + F^0(\tau,\theta,\phi) e_0^{\phantom{I}a}(\tau)\right] +\mathcal{O}\left(\frac{r^4}{L^3}\right),
\end{equation}
where, as before, $x^\alpha = (x^0,x^1,x^i) = (\tau,r,\theta,\phi)$ are coordinates adapted to the observers: the angular coordinates $x^i=(\theta,\phi)$ are parameters that label the observers' worldlines; the radial coordinate, $r$, parameterizes the size of the RQF; and the time coordinate, $\tau$, represents our choice of simultaneities on $\mathcal B$. The three arbitrary functions $F^I(\tau,\theta,\phi)$ allow us to perturb the observers' worldlines to satisfy the three conditions $\sigma_{ij} = r^2\mathbb{S}_{ij}$ needed to achieve a round sphere RQF of areal radius $r$. (An RQF of generic shape and size need only satisfy the three weaker conditions, $\partial \sigma_{ij}/\partial\tau = 0$.) The arbitrary function $F^0(\tau,\theta,\phi)$ allows us to perturb the choice of simultaneities on $\mathcal B$ to achieve other natural geometrical conditions, which will be discussed later. For the time being, we will keep the four functions $F^I$ and $F^0$ {\it arbitrary}.

Choosing $\psi^a=\frac{1}{c}u^a$ in equation~(\ref{QuasilocalIntegrated}) makes this a quasilocal energy conservation equation. Using GRTensorII running under Maple, we compute the individual terms in equation~(\ref{QuasilocalIntegrated}) and find:
\begin{align}
\frac{1}{c}\, d\mathcal{S}\,  T_{\mathcal B}^{ab} u^{\mathcal{S}}_a \psi_b &=
r^2\,d\mathbb{S}\,\left\{ -\frac{c^4}{4\pi G}\frac{1}{r} + \gamma^2\frac{E^2}{16\pi}\sin^2\theta (5\cos^2\theta+1)\,r +  (C + \Phi_F + \Psi_F)\, r + \frac{c^4}{GL} \times \mathcal{O}\left(\frac{r^2}{L^2}\right)  \right\},\label{RQFEnergyDensity}\\
d \mathcal{B} \,  T^{ab} n_a \psi_b &= c\,d\tau\,r^2\,d\mathbb{S}\, \beta\gamma^2\frac{E^2}{4\pi}\cos\theta\, \left\{ 1 + \frac{1}{c^2}\mathtt{a}(\tau)r\cos\theta + \mathcal{O}\left(\frac{r^2}{L^2}\right)  \right\},\label{RQFPoyntingDensity}\\
d \mathcal{B} \,  T_{\mathcal B}^{ab} \left( D_{(a} \psi_{b)} \right) &= d \mathcal{B}\, \left( \frac{1}{c}\,\alpha_a \mathcal{P}^a +  \mathcal{S}^{ab} \theta_{ab} \right) =
-c\,d\tau\,r^2\,d\mathbb{S}\, \beta\gamma^2\frac{E^2}{8\pi}\sin^2 \theta\frac{\mathtt{a}(\tau)r}{c^2}\, \left\{ 1 + \mathcal{O}\left(\frac{r}{L}\right)  \right\}
+ d \mathcal{B}\,\mathcal{S}^{ab} \theta_{ab},\label{RQFGravityDensity}
\end{align}
where $d\mathbb{S} = \sin\theta \,d\theta \,d\phi$ is the area element on the unit round sphere, as before, and $\gamma(\tau)$ and $\beta(\tau)$ are the same Lorentz transformation parameters we defined after equation~(\ref{BulkEM2}). We have also inserted the relation $L = c^2/\sqrt{G} E$ where appropriate. Let us analyze each of these terms separately.

Equation~(\ref{RQFEnergyDensity}) represents the quasilocal energy density. The first term on the right hand side, at order $1/r$ inside the braces, is a vacuum energy. Insofar as it is constant in time, this term does not contribute to changes in the energy inside the sphere, and so can be ignored. The quantity $C$ is a function of the angles $(\theta,\phi)$, but not $\tau$. So for the same reason just cited, it can also be ignored. The quantity $\Phi_F$ is a homogeneous function of the angular derivatives of the $F^I$, which is a total derivative. It integrates to zero for {\it any} choice of the arbitrary functions $F^I$, and so can also be ignored. The quantity $\Psi_F$ is an inhomogeneous function of the angular derivatives of the $F^I$, which is a linear combination of the three round sphere RQF conditions, $\sigma_{ij} - r^2\mathbb{S}_{ij}=0$. It is zero when we choose a set of functions $F^I$ that satisfy these conditions (which we will do later), and so can also be ignored. The only nontrivial term is the second one. Insofar as only the integrated quasilocal energy density is physically meaningful, we can subtract off the spherical harmonics that integrate to zero, leaving the {\it effective} quasilocal energy density:
\begin{equation}
\left\{\frac{1}{c}\, d\mathcal{S}\,  T_{\mathcal B}^{ab} u^{\mathcal{S}}_a \psi_b \right\}_{\rm effective}=
r^2\,d\mathbb{S}\,\left\{  \gamma^2\frac{E^2}{12\pi}\,r + \frac{c^4}{GL} \times \mathcal{O}\left(\frac{r^2}{L^2}\right)  \right\}.\label{EffectiveRQFEnergyDensity}
\end{equation}

Equation~(\ref{RQFPoyntingDensity}) represents the quasilocal Poynting flux. The quantity in braces is (at least up to the order indicated) equal to the lapse function, $N(x)=1 + \frac{1}{c^2}\mathtt{a}(\tau)r\cos\theta + \mathcal{O}\left(\frac{r^2}{L^2}\right)$, which is contained in $d{\mathcal B}$. Up to the lowest nontrivial order in $r$, this lapse function agrees with that found in \S\ref{Local}; as before, it is this non-isotropic time dilation that allows for a nonzero accumulation of Poynting flux in the sphere. Note that, up to the order indicated, the Poynting flux is independent of our choice of arbitrary functions $F^I$ and $F^0$. Equation~(\ref{RQFGravityDensity}) represents the sum of the quasilocal gravitational flux and the quasilocal ``stress times strain" term---recall equation~(\ref{QuasilocalTransfer}). When we impose the RQF rigidity conditions on the functions $F^I$, the latter term vanishes (i.e., the strain rate tensor, $\theta_{ab}$, vanishes). This leaves only the quasilocal gravitational flux term, $\alpha_a \mathcal{P}^a$, to consider. Note that, up to the order indicated, the gravitational flux is also independent of our choice of arbitrary functions $F^I$ and $F^0$.

Putting these results together, with the assumption that the RQF rigidity conditions can be satisfied (we'll demonstrate this below), the quasilocal energy conservation law in equation~(\ref{QuasilocalIntegrated}) reads, at order $r^3$,
\begin{equation} \label{QuasilocalIntegratedParadox}
\int\limits_{\mathcal{S}_f - \mathcal{S}_i}  r^2\,d\mathbb{S}\,\gamma^2\frac{E^2}{12\pi}\,r
= \int\limits_{\mathcal{B}}  c\,d\tau\,r^2\,d\mathbb{S}\,\left\{
\beta\gamma^2\frac{E^2}{4\pi}\cos\theta\, \left( 1 + \frac{1}{c^2}\mathtt{a}(\tau)r\cos\theta \right)
+\beta\gamma^2\frac{E^2}{8\pi}\sin^2 \theta\frac{\mathtt{a}(\tau)r}{c^2}
\right\}.
\end{equation}
This quasilocal conservation law is to be compared with the local conservation law in equation~(\ref{ParadoxConservationLaw}). As we did for the latter equation, we will take the time interval to be infinitesimal: $\tau_f -\tau_i = \Delta\tau$. Using the relation $\frac{d}{d\tau}  \gamma^2   = 2 \beta \gamma^2\frac{\mathtt{a}}{c} $ [compare this with $\frac{d}{d\tau} \left( (1 + \beta^2) \gamma^2  \right) = 4 \beta \gamma^2\frac{\mathtt{a}}{c} $, which differs by a factor of 2], the left hand side of equation~(\ref{QuasilocalIntegratedParadox}) then integrates to precisely the same change in electromagnetic energy, $\Delta E$, given by the local conservation law, and also equation~(\ref{E(1)}) (with the cylinder volume, $V=AL$, replaced with the sphere volume, $V=\frac{4}{3} \pi r^3$). The quasilocal Poynting flux integral over $\mathcal{B}$ on the right hand side of equation~(\ref{QuasilocalIntegratedParadox}) is, as mentioned earlier, identical to the local Poynting flux integral in equation~(\ref{ParadoxConservationLaw}), and so yields the same change in electromagnetic energy, $\Delta E_{\rm Poynting}$, as found in that case, and also given in equation~(\ref{E(2)}) (with, again, the cylinder volume replaced with the sphere volume). We emphasize once more the role of the non-isotropic time dilation (the lapse function in parentheses) that allows for a nonzero accumulation of Poynting flux.

The quasilocal gravitational flux integral over $\mathcal{B}$ on the far right hand side of equation~(\ref{QuasilocalIntegratedParadox}) is, as discussed earlier, the surface flux version of the bulk ``acceleration times momentum" energy transfer term in the local conservation law. The difference is that the local momentum volume density, $\mathbb{P}^a$, has been replaced with the quasilocal momentum surface density, ${\mathcal P}^a$, and the integration is not over a volume, but the boundary of that volume. We stress again that the actual mechanism of the energy transfer is a general relativistic effect, viz., the vector cross product between the boundary observers' acceleration and the precession rate of their gyroscopes due to the frame dragging caused by the electromagnetic momentum inside the system. All together, then, half of the energy flux is due to the electromagnetic field (with the flux entering mainly near the poles of the sphere, i.e., $\cos^2\theta$), and the other half is due to the gravitational field (with the flux entering mainly near the equator of the sphere, i.e., $\sin^2\theta$). Both contribute to a changing electromagnetic energy on the left hand side of the quasilocal conservation law.

One might wonder how a flux of {\it gravitational} energy through the boundary of the system becomes {\it electromagnetic} energy inside the system. It seems that in general relativity there is no distinction between the two forms of energy. All forms of energy are equivalent. This is not surprising when we look at the metric in equation~(\ref{ElectricSpacetime}). The geometry is nontrivial and so presumably contains gravitational energy (in some nonlocal, or quasilocal sense), and the nontrivial geometry in turn is intimately connected to the electric field. They cannot be disentangled.

All that remains in this section is to verify that we actually {\it can} solve the RQF rigidity conditions, viz., the three differential constraints $\sigma_{ij} = r^2\mathbb{S}_{ij}$ on the three functions $F^I(\tau,\theta,\phi)$ which ensure that the observers' frame is a rigid round sphere of areal radius $r$. As argued in reference \cite{EMM2011}, such RQF solutions always exist in a generic spacetime; moreover, they are unique up to motions of the RQF equivalent to those of rigid motion in Newtonian space-time, viz., six arbitrary time-dependent degrees of freedom: three translations and three rotations. Aiming for the simplest solution, we first set $F^1(\tau,\theta,\phi)$ and $F^2(\tau,\theta,\phi)$ to zero, which eliminates translations of the RQF away from ${\mathcal C}_0$ in the $X$ and $Y$ directions. We next demand that the RQF is not rotating, i.e., that the twist of the observers' congruence is zero, i.e., that the observers' four-velocity, $u^a$, is hypersurface orthogonal (as a vector field in $\mathcal B$). This means that, by adjusting $F^0(\tau,\theta,\phi)$ appropriately, we can always find a time foliation of $\mathcal B$ (i.e., choose the observers' surfaces of simultaneity) such that $u^a$ is orthogonal to them, which is equivalent to the two conditions $u_i=0$. Using GRTensorII we can readily find a solution that satisfies all of these conditions, given by:
\begin{align}
F^3(\tau,\theta,\phi) &= \frac{1}{6}\,\gamma^2\,\cos\theta(3-\cos^2\theta)
-\frac{1}{6}\cos\theta(\cos^2\theta+3\sin^2\theta\cos^2\phi),\label{F3}\\
F^0(\tau,\theta,\phi) &= -\frac{1}{3}\,\beta\gamma^2\,\cos\theta(3-\cos^2\theta).\label{F0}
\end{align}
Note that the $\phi$ dependence in $F^3(\tau,\theta,\phi)$ is not surprising since the electric field, being parallel to the $\phi=0$ plane, breaks the azimuthal symmetry---the electric field affects the geometry, and hence the coordinate embedding of the RQF. One can show that these results are in agreement with the more generally derived formulas found in reference \cite{EMM2011}.

\section{Summary and Conclusions}\label{Conclusions}

Using the standard, local way of constructing conservation laws---see equations~(\ref{BulkDifferential}) and (\ref{BulkIntegrated})---we analyzed conservation of energy in the context of a simple example in special relativity, viz., a box, rigidly accelerating along the $Z$-axis, that is immersed in a transverse, uniform electric field. According to the local energy conservation law, the electromagnetic energy inside the box increases due to two separate mechanisms: (1) Half of the increasing energy is due to energy flowing in from outside the box via a Poynting flux. Interestingly, even though the instantaneous proper Poynting vector is uniform throughout the box (suggesting no net flux), there is a net proper time-integrated flux due to the acceleration-induced relative time dilation between observers at the top and bottom of the box. (2) The other half of the increasing energy is {\it not} due to energy flowing in from the outside; rather, it is a bulk effect due to the co-moving observers accelerating relative to the existing electromagnetic momentum inside the box, i.e., an ``acceleration times momentum" energy transfer mechanism familiar from classical mechanics, integrated over the volume of the box.

One might wonder if these two energy transfer mechanisms adequately explain what's happening. The answer is: No. First, in special relativity we assume that spacetime is flat, even though there is an electromagnetic field present. This precludes any possible general relativistic effects such as a flux of gravitational energy associated with the curvature of the spacetime caused by the electromagnetic field. Secondly, even in the context of general relativity, a {\it local} conservation law cannot properly capture all of the gravitational physics; for example, gravitational energy is not localizable---there is no such thing as a gravitational energy per unit volume, so a local conservation law cannot tell the whole story. Another way to see this problem is to notice that the standard local conservation law is {\it homogeneous} in the matter stress-energy-momentum tensor, and so has nothing to say in a matter-free region of space, even though that region may contain dynamical curvature, e.g., gravitational waves.

To address these shortcomings in general, and in particular see what's really happening with regards to the increasing energy inside the box, we constructed a {\it quasi}local conservation law based on the Brown \& York quasilocal {\it total} (matter plus gravitational) stress-energy-momentum tensor defined on the history of the boundary of a spatial volume---see equations~(\ref{QuasilocalDifferential}) and (\ref{QuasilocalIntegrated}). Using the energy form of this quasilocal conservation law, we analyzed the general relativistic analogue of the simple example described above. We found, again, that the electromagnetic energy inside the box increases due to two separate mechanisms. The first mechanism is identical in form to the Poynting flux mechanism described above [number (1)], but conceptually it has a completely different origin: it is actually analogous to a surface flux version of the $\vec{j}\cdot\vec{E}$ bulk term in the standard Poynting theorem, except instead of energy being transferred locally from the electromagnetic field to the four-current source of the field, it represents energy being transferred quasilocally from the matter fields to the ``system" contained inside the box. The second mechanism is, at first sight, conceptually identical to the ``acceleration times momentum" energy transfer mechanism described above [number (2)], except the volume integral has been converted to a {\it surface flux} integral over the boundary of the box. Going further, we argued that this surface flux is actually a {\it gravitational} energy flux exactly analogous to the electromagnetic Poynting flux, $\vec{E}\times\vec{B}$, with $\vec{E}$ replaced by the acceleration of the co-moving observers on the boundary of the box, and $\vec{B}$ replaced by the precession rate of their gyroscopes due to the frame dragging caused by the electromagnetic momentum (in general, matter and gravitational momentum) flowing through the box. In {\it both} cases, now, energy is entering from outside the box via surface fluxes: half is an electromagnetic energy flux (entering the box through its top and bottom) and the other half is a gravitational energy flux (entering the box through its {\it sides}). Because an electromagnetic field is intimately intertwined with the spacetime curvature it produces, general relativity does not distinguish between electromagnetic and gravitational energy entering the box---both contribute on the same footing to the increasing electromagnetic energy inside the box.

We can understand the second, gravitational energy flux mechanism intuitively as follows. Imagine being inside an accelerating box in empty space, which contains a freely-floating massive body that appears to be accelerating toward you. From your perspective, its kinetic energy is increasing. Where does the increasing kinetic energy come from? We could ``explain" it as simply the ``acceleration times momentum" energy transfer mechanism familiar from classical mechanics. Alternatively, we could invoke the equivalence principle and say that our box is at rest in a uniform gravitational field, and that the kinetic energy is coming from the gravitational field outside the box via some kind of gravitational energy flux passing through the boundary of the box. An analogous mechanism exists in the context of electrodynamics that explains, in detail, how an accelerating charge acquires kinetic energy from the external electric field causing the particle's acceleration \cite{Gaidukov}; moreover, the gravitational version of this mechanism has been shown to exist in the context of the linearized, weak field approximation to general relativity \cite{Krumm,Matos}. In this paper we have established the existence of this very basic, but subtle mechanism in the full, nonlinear general theory of relativity, with arbitrary matter and no approximations. Our analysis made use of the concept of {\it Rigid Quasilocal Frames} (RQFs), discussed more fully in references \cite{EMM2009, EMM2011}, in order to properly isolate the relevant energy fluxes passing through the boundary of the box.

In conclusion, we have explained the bulk ``acceleration times momentum" energy transfer mechanism familiar in classical mechanics {\it exactly} in terms of a simple, operationally defined, purely geometrical, general relativistic gravitational energy flux passing through the {\it boundary} of the volume in question. Naively, one might argue that since there is no $G$ or $c$ in ``acceleration times momentum", this cannot be a general relativistic effect. But it {\it is}. It is based on frame dragging (the gravitational analogue of ``$\vec{B}$" in ``$\vec{E}\times\vec{B}$"), which is a general relativistic effect. We don't notice this mechanism in our day-to-day experiences because the typical gyroscopic precession rate vector due to a nearby object in motion is very tiny; but it is precisely this vector, multiplied by the huge number $\frac{c^2}{8\pi G}$, that we identify as the ``momentum" of the object [more precisely, the quasilocal momentum surface density, rotated by $90$ degrees---see equation~(\ref{GravitationalPoynting})]. This general relativistic gravitational energy flux mechanism is what's {\it really} happening in the bulk ``acceleration times momentum" energy transfer mechanism in classical mechanics. This deeper understanding would not be possible in the context of local conservation laws, and is a nice example of why we need {\it quasi}local conservation laws.

\section*{Acknowledgments}

This work was supported in part by the Natural Sciences and Engineering Research Council of Canada.


\begin{thebibliography}{99}
\bibitem{MTW} C.~W.~Misner, K.~S.~Thorne and J.~A.~Wheeler, {\it Gravitation}, ISBN 0-7167-0344-0 (W.~H.~Freeman and Company, 1973).
\bibitem{BY1993} J.~D.~Brown and J.~W.~York, Phys. Rev. D \textbf{47}, 4, 1407-1419 (1993).
\bibitem{EMM2009} R.~J.~Epp, R.~B.~Mann and P.~L.~McGrath, Class. Quant. Grav. \textbf{26}, 035015 (2009).
\bibitem{EMM2011} R.~J.~Epp, R.~B.~Mann and P.~L.~McGrath, {\it Classical and Quantum Gravity: Theory, Analysis, and Applications}, ISBN 978-1-61122-957-8 (Nova Science Publishers, 2012), ch. 14.
\bibitem{Bell} J.~S.~Bell, {\it Speakable and Unspeakable in Quantum Mechanics}, ISBN 0-521-33495-0 (Cambridge University Press, 1987), pp. 67.
\bibitem{Gaidukov} G.~N.~Gaidukov and A.~A.~Abramov, Phys. Usp. \textbf{51}, 163 (2008).
\bibitem{Mashhoon} B.~Mashhoon, {\it The Measurement of Gravitomagnetism: A Challenging Enterprise}, ISBN 978-1-60021-002-0 (Nova Science Publishers, 2007), pp. 29.
\bibitem{Krumm} P.~Krumm and D.~Bedford, Am.~.J.~Phys. \textbf{55}, 362 (1987).
\bibitem{Matos} C.~J.~de~Matos and M.~Tajmar, e-Print: gr-qc/0107014 (2001).
\bibitem{Bertotti} B.~Bertotti, Phys. Rev. \textbf{116}, 1331 (1959).
\bibitem{Stephani} H.~Stephani, D.~Kramer, M.~MacCallum, C.~Hoenselaers, and E.~Herlt, {\it Exact Solutions of Einstein's Field Equations}, ISBN 0-521-46136-7 (Cambridge University Press, 2003), pp. 176.
\end{thebibliography}
\end{document}